# An improved design of an inductive fault current limiter based on a superconducting cylinder

V. Sokolovsky[1], V. Meerovich[1], L. I. Chubraeva[2], I. Vajda[3]

[1]Physics Department, Ben-Gurion University, Beer-Sheva 84105, Israel

[2]Department of Control Systems and Nano-technologies, Saint-Petersburg University of Aerospace Instrumentation, Saint-Petersburg, Russia

[3]Department of Electric Power Engineering, Budapest University of Technology and Economics, Egry Jozsef utca 18, H-111 Budapest, Hungary

E-mail: **victorm@bgu.ac.il**

**Abstract**

The paper deals with basic designs of a fault current limiter of the transformer type which differ each other by the mutual location of a primary winding and a superconducting short-circuited cylinder. Theoretical study of the main parameters of the different designs is performed in the framework of the critical state model and shows that the most effective is a design in which the primary winding is divided to two sections with equal turn numbers. The sections are placed inside and outside of the cylinder and connected in series. Such arrangement of the windings leads to a substantial reduction of AC losses in the superconducting cylinder, an increase of the activation current and a decrease of the inductive reactance in the normal regime of a protected circuit. The experimental results obtained on the laboratory model with a BSSCO cylinder confirm the theoretical predictions.

(Some figures in this article are in color only in the electronic version)

1. **Introduction**

A superconducting (SC) short-circuited winding is used as a switching element of various current limiting devices of a transformer type [1-7]. One widely studied design is an inductive fault current limiter (FCL) consisting of two magnetically coupled windings: a primary winding, which is inserted in series in a power circuit, and a secondary SC short-circuited winding [1-4]. The S-N transition in the SC winding causes the increase of the resistance of this winding followed by the increase of the FCL impedance.

Numerous studies of the inductive FCL deal with two different arrangements:
   a) the SC winding is coaxially placed inside the primary winding (so called magnetic shielding design) [1,3];
   b) the primary winding is coaxially placed inside the SC winding [2,4].

If the SC winding is fabricated from a well-stabilized SC wire and short-circuited by a special SC switching element [2], the magnetic flux penetrates its body like in the case of non-superconducting winding of a transformer.

Therefore, the relative positions of the windings do not influence on the characteristics of the device and the arrangement should be chosen from considerations of the best cryostat design, required electrical isolation and etc.

Another situation is observed if a SC winding is formed by SC hollow cylinders. The currents induced in the cylinder not only compensate the magnetic flux in the hollow but also shield a part of the cylinder from the magnetic field. The question arises how the mutual bracing of the cylinders and windings affects the main characteristics of the device: impedance in the normal and fault regimes, AC losses in the SC cylinder, activation current, and time of the recovery of the initial state after a fault event. What arrangement is better from point of view of the most effective operation of a FCL?

The purpose of this paper is to compare the characteristics of the FCL for different mutual locations of the SC cylinder and primary winding. It is shown that the most effective design is achieved by dividing the primary winding into two sections. This result is generalized to more complicated transformer devices employing SC cylinders or rings.

2. **Theoretical consideration**

*2.1. FCL impedance*

The sketches of Figs. 1a and 1b show two mutual locations of the SC cylinder and primary winding in a fault current limiter with an yokeless magnetic system [1-4].

For calculations, we will use the approximation of infinite long windings and assume that: 1) all the magnetic flux lines are straight and parallel to the $z$-axis, 2) the magnetic system is linear, 3) the winding and cylinder thicknesses and all the gaps much less than the diameter of the ferromagnetic core. Under these assumptions, usual formulas for transformers and electrical reactors can be used for the calculation of the inductance of the FCL under the normal and fault regimes [2].

Under the normal regime of the protected circuit the current in a SC cylinder compensates the magnetic flux in the interior space of the cylinder. The device inductance $L_s$ is the leakage inductance of a transformer with a short-circuited secondary winding and it depends strongly on the thicknesses of the winding $b_c$ and cylinder wall $\Delta$ and on the gap between them $\delta$ [2, 8]:

$$L_s = \mu_0 W^2 \pi D_g \left(b_c/3 + \delta + \Delta/3\right)/h_c, \tag{1}$$

where $W$ and $h_c$ are the turn number and height of the primary winding, respectively, $D_g$ is the middle diameter of the gap.

Under the assumption that in the fault regime the cylinder is heated above the critical temperature and passes into the normal state with a large resistance, the inductance in this regime $L_n$ can be estimated as the inductance of the device without a cylinder (one-winding electrical reactor) [2]. The value of the inductance depends on the design of the magnetic system. Here we consider the design with an open magnetic system without yoke. This is the most effective and widely applied design of an inductive fault current limiter [1-4]. As was shown in [8], the inductance for this design can be evaluated using a simple expression:

$$L_n = \mu_0 W^2 \pi D k_1 / 1.76 \tag{2}$$

where $D$ is the diameter of the ferromagnetic core, $k_1$ is a geometric coefficient close to unity.

For real devices $D_g \approx D$, which is determined by the voltage class of the protected circuit. Expressions (1) and (2) show that the values of the FCL impedance under both regimes do not depend on the mutual location of the cylinder and primary winding.

Let us consider the case, when the primary current amplitude exceeds the value $I_c/W$ ($I_c$ is the critical current determined as $J_c \Delta h_{cyl}$, where $J_c$ is the critical current density and $h_{cyl}$ is the cylinder height) and the magnetic flux penetrates through the SC cylinder into the magnetic core. The increase of the primary current above $I_c/W$ does not change the current in the cylinder which equals the critical value; this value can only decrease due to heating. The profiles of the magnetic field for this case are shown by the lines 2 in Fig. 1. For real devices, the core diameter is much larger than the cylinder wall thickness and all the gaps, so that the voltage drop $u$ across the primary winding is practically determined by the magnetic flux in the core and can be calculated using the equations for a two-winding transformer where the secondary current equals the critical value:

$$u = r_w i + L_n \frac{di}{dt} - M \frac{sign(i)}{W} \frac{dI_c}{dt} \qquad \text{at } |i| > I_c/w \tag{3}$$

where $i$ is the instantaneous primary current, $r_w$ is the resistance of the primary winding, $M$ is the mutual inductance of the primary winding and cylinder calculated as $M = L_n \sqrt{1 - L_s/L_n}$. Sign "-" appears because the cylinder current is opposite to the primary one, while $I_c$ is taken as a positive value.

Eq. (3) determines behavior of the device under transient process. If heating is negligible, $I_c$ = const and the voltage across the winding is determined by the impedance given by (2). If heating is noticeable, the analysis of the process is based on the simultaneous solution of the equation describing of the thermal state of the superconductor and the equations of the electric circuit [6,9-11]. The time which is required for heating of a superconductor up to the critical temperature is determined by the dissipated energy and thermal capacity of the superconductor.

*2.2. Losses in the SC cylinder under normal and fault regimes*

a) If the SC cylinder is located inside the primary winding, the magnetic field penetrates the cylinder wall from outer surface, at $r = R_{out}$ (Fig. 1a). The Bean model gives the maximum depth $d_m$ of the field penetration[1]

$$d_m = H_0 / J_c , \qquad (4)$$

where $H_0$ is the amplitude of the magnetic field at the cylinder surface which is determined as

$$H_0 = w I_0, \qquad (5)$$

here $w$ is the number of turns per unit of the winding length, $I_0$ is the amplitude of the current in the winding.

At the penetration depth $d_m$ less than the wall thickness $\Delta$, i.e. when the magnetic field amplitude less than the complete penetration field $H_p = J_c \Delta$, the profile of the field is shown by the line 1 in Fig. 1a. Magnetic field is zero at $r < R_{in}$ where $R_{in}$ is the inner radius of the cylinder.

The Bean model gives the following expression for the AC loss density per period and per length unit of the SC cylinder [9]:

$$p_{out} = 2\pi R_{out} \frac{2\mu_0}{3 J_c} H_0^3 \left( 1 - \frac{H_0}{2 J_c R_{out}} \right) . \qquad (6)$$

Eq. (6) is valid when $d_m \leq \Delta$. This case corresponds to the normal regime of the protected circuit.

b) If a cylinder is placed outside the winding (Fig. 1b), the magnetic field produced by the winding current is zero at the cylinder under any regime. The magnetic field in the cylinder is produced only by the current induced in the cylinder according the Faraday law. In this case, the magnetic field penetrates into the cylinder wall from the inside, at $r = R_{in}$, and it is always zero at the outer surface, at $r = R_{out}$. Assuming that cooling and transition of the cylinder into the superconducting state occurs at zero current in the winding, under normal regimes of the protected circuit the cylinder current can be determined from the condition that the total magnetic flux in the cross-section of the SC cylinder is zero. The cross-section of the cylinder is larger than that of the winding. Therefore, the magnetic flux in the core is not zero even under the normal regime as opposed to the case when cylinder is inside the winding. The profile of the magnetic field is shown by the line 1 in Fig. 1b. The current in the cylinder per unit of the length is

$$i_{cyl} = -k_b w i(t), \qquad (7)$$

where $k_b$ is the coupling coefficient of the cylinder and primary winding. This coefficient can be estimated as the ratio of the total magnetic flux to the flux through the cross-section of the primary winding. The estimation gives that $k_b$ can be taken equal to unity for large-scale devices.

The Bean model gives the following AC loss density per a period and per a length unit of the SC cylinder ($d_m \leq \Delta$) [9]:

---

[1] In this paper we assume that the cylinder and the winding have the same height. The obtained results can be generalized for the case of various lengths. For example, the magnetic field at the cylinder surface is determined as $H_0 = w I_0 h_{cyl} / h_c$.

$$p_{in} = 2\pi R_{in} \frac{2\mu_0}{3J_c} k_b^3 H_0^3 \left(1 + \frac{H_0 k_b}{2 J_c R_{in}}\right). \tag{8}$$

At $k_b \cong 1$ and $R_{out} \cong R_{in}$, Eq. (13) differs from (6) only by the sign of the second term in the brackets. The maximum of this term is the ratio of the thickness of the cylinder wall $\Delta$ to the cylinder diameter (internal or external) and much less than unity.

c) Let us compare the losses in two designs in the case when $d_m > \Delta$, i.e. the magnetic field penetrates through the cylinder wall. This case corresponds to the fault regime; the profiles of the magnetic field exceeding the complete penetration field are shown by the lines 2 in Fig. 1. The losses per unit of the superconductor volume are $EJ_c(T)$ and for comparison it is enough to compare the electric field induced in the superconductor. The electric field is determined as a solution of the following Maxwell equations:

$$\frac{\partial H}{\partial r} = -J_c; \tag{9}$$

$$\frac{1}{r}\frac{\partial (rE)}{\partial r} = -\mu_0 \frac{\partial H}{\partial t}. \tag{10}$$

In the design of Fig.1a the magnetic field at the outer surface of the cylinder, at $r = R_{out}$, is determined by only the field produced by the current in the primary winding: $H|_{r=R_{out}} = wi(t)$. Assuming that the temperature is uniform over the cylinder and $|i| > I_c/W$, the solution of Eq. (9) is presented as:

$$H = \begin{cases} wi(t) - sign(i) J_c(T)(R_{out} - r) & \text{if } r > R_{in} \\ wi(t) - sign(i) J_c(T) \Delta & \text{if } r \leq R_{in} \end{cases}. \tag{11}$$

At the total penetration, the electric field in the cylinder wall can be presented as a sum of three parts: the electric field induced by variation of the external magnetic field in the wall ($w di(t)/dt$), the electric field caused by change of the critical current density due to heating, and the electric field $E_0$ induced at the inner surface of the cylinder, at $r = R_{in}$, by variation of the magnetic flux in the cylinder hollow. The solution obtained from Eq. (10) has the form

$$E = -\frac{\mu_0}{r}\left\{\frac{w}{2}\frac{di(t)}{dt}(r^2 - R_{in}^2) - sign(i)\frac{dJ_c(T)}{dt}\left[\frac{R_{out}}{2}(r^2 - R_{in}^2) + \frac{r^3 - R_{in}^3}{3}\right]\right\} + E_0 \tag{12}$$

where $E_0$ is practically determined by variation of the magnetic flux in the ferromagnetic core:

$$E_0 = -\mu_0 \mu_r \frac{D^2}{8R_{in}}\frac{d}{dt}[wi - sign(i) J_c \Delta] = -\frac{L_n}{2\pi R_{in} W}\frac{d(i - sign(i) I_c /W)}{dt}, \tag{13}$$

where $\mu_r$ is the relative permeability of the ferromagnetic core.

The losses are determined from Eqs. (12) and (13) by the integration over the cylinder volume.

The first term (in brackets) in expression (12) for $E$ is small because $(E - E_0)/E_0 \sim \frac{\Delta}{\mu_r R_{in}}$. Therefore, the losses can be evaluated using $E_0$ only given by Eq. (13).

If the SC cylinder is located outside the primary winding (Fig. 1b), the electric field is also approximately determined by Eq. (13).

Thus, all the differences in the parameters of two designs of an inductive FCL are determined by ratios of the winding and/or cylinder thicknesses and their diameters and/or by the ratio of the effective gap to the core diameter. When small laboratory models of the device are studied, the wall thickness and diameter of the cylinder can be comparable resulting in sufficient difference of the parameters. In real power devices all the ratios mentioned above are much less than unity. As a result, the main parameters of the two considered designs do not differ essentially. The choice of a design is determined by the cryostat design and by the gaps required for providing electrical isolation.

*2.3. Two-sectional primary coil*

The obtained results allow us to propose the design of an inductive FCL with a two-sectional primary winding: the first section is placed inside the cylinder while the second – outside (Fig.1c). The sections have the same turn number and connected in series. The device impedance in the limitation regime, when the cylinder is in the normal state, is practically equaled to the impedance of the device with a single winding having the turn number equaled to the sum of turn numbers in both sections. The advantages of dividing the winding are

1) Reduction of AC losses in the superconducting cylinder under the normal regime of the protected circuit. The magnetic field at each side of the surface is a half of the magnetic field produced by the single winding and, according Eqs. (6) and (13), the total losses are four times less than in a design with a single winding.

2) Increase of the activation current due to two factors: decrease of AC losses and decrease of the external magnetic field. AC losses can increase the temperature inside a bulk superconductor by several Kelvins [12, 13] and, hence, decrease the critical current. In the framework of the Bean model the partition of the winding into two parts does not change the total penetration depth of the magnetic field. In reality, the critical current density depends on a local magnetic field. For example, the Kim-Anderson model gives

$$J_c(H) = J_c(0) H_b / (H_b + H), \tag{14}$$

where $H_b$ is the constant of fitting.

Integrating Eq. (9) gives the maximum penetration depth as a function of the amplitude of the magnetic field applied to one side:

$$d_m = \frac{H_0}{J_c} + \frac{H_0^2}{2 H_b J_c}. \tag{15}$$

The first term in (15) is the maximum penetration depth given by the Bean model, the second term is connected with the dependence of the critical current density on a local magnetic field. Fig. 2 presents the ratio, $k_{dm} = d_m(H_0)/2d_m(H_0/2)$, of the maximum penetration depth in the device with a single primary winding and in the device with a two-sectional winding. One can see that at $H_0 \gg H_b$ the ratio tends to 2, $k_{dm} \to 2$, i.e. the activation current can be doubled.

3) Decrease of the inductance $L_s$ in the normal regime and increase the ratio of the device impedances under the normal and limitation regime $L_n/L_s$. The SC cylinder separates the magnetic fluxes of both sections of the primary winding (Fig. 1c), so that the total impedance is determined by the sum of the impedances of the sections, each of which has a half of the turns. The inductance of each section becomes four times less than the inductance of the single winding in previous designs. Therefore, the total inductance of a two-sectional winding in the normal regime is two times less than the inductance of a single winding.

## 3.    Experimental investigations

*3.1. Experimental model*
The experimental model of an inductive FCL constitutes an open core transformer and it is similar to the models previously investigated by us [14]. The main difference from the previously investigated models is that the primary winding of the model consists of two parts. Each of the parts is a 400-turn coil. The first of them is inserted into a superconducting melt cast 2212 BSCCO cylinder; the second coil is placed coaxially outside the cylinder. The coils and cylinder are centered on a 10×12 mm$^2$ cross- section ferromagnetic core. The model was immersed in liquid nitrogen at the temperature of 77 K.

The inner coil with the outer diameter of 20 mm and thickness of 1 mm was made out of 0.35 mm cooper wire and has the resistance of 0.45 Ω at 77 K. The outer coil with the outer diameter of 50 mm and thickness of about 2 mm has the resistance of 0.72 Ω at 77 K and was made out of 0.5 mm cooper wire. The height of both coils was 38 mm.

The tested BSCCO cylinder has 31 mm in height, 35 mm in outside diameter, and 2 mm in wall thickness. It was cut out from the cylinder with the wall thickness of 5 mm fabricated using the melt cast technology [15,16]. The characterization of the specimens fabricated with the same technology gave the critical current density of about 570 A/cm$^2$ at 77 K (determined by DC four-point method from 1 µV/cm criterion) and the critical temperature of 94 K [15]. Up to the electric field intensity of $3 \cdot 10^{-3}$ V/cm, the E-J characteristic is well fitted by the power law with the exponent equaled to 7-8.

Our experimental setup for FCL testing has been described in details in [14]. The testing was carried out in the following stages: first we measured the parameters of the model without a superconducting cylinder. Then the voltage-current characteristic of the model and losses in the cylinder under steady-state regime were measured in the process of step by step increasing the current in the coils. At every current step we waited several minutes to achieve thermal equilibrium of the cylinder. At the last stage of the research, the model response to a sudden fault was investigated. The experiments were carried out for each coil separately and for the case when the coils were connected in series. The last allowed us to compare a device with a single winding and a device with the winding parted into two sections.

*3.2. Experimental results and their discussion*
The inductive impedance of the coils in the model without a superconducting cylinder: the inner coil is 4.8 Ω, the outer – 5.1 Ω, and when the coils are connected in series this impedance is about 18.8 Ω. In spite of a relative big gap between the coils, the inductive impedances differ only by about 6%. When the coils are connected in series, the impedance is about 4 times larger than the impedance of each coil.

The voltage-current characteristics of the model are presented in Fig. 3. Here the last measured points correspond to the maximum non-quenching current. The increase of a current above this value only by ~0.05 A leads to quenching and increase of the model impedance up to a value very close to the impedance without cylinder.

At low currents the voltage-current characteristics are well fitted by linear dependences. The reactances of the coils calculated from measurements of the impedance and resistance: 0.84 Ω – inner coil, 1.14 Ω - outer coil, and 0.55 Ω – two coils connected in series (the last reactance is reduced to the total number of turns equaled to 400). The 25% difference in the reactance of two coils is explained by the difference in the middle diameters and thicknesses of the gaps between the coils and cylinder.

The total reactance of two coils becomes by ~ 10% higher than the expected value equaled to a half of the sum of the coil reactances. The reason of this is some magnetic coupling of the coils due to incomplete shielding the magnetic flux by the superconducting cylinder of a finite length. When the cylinder is in the SC state, the measurement shows that the voltage across the inner coil was only 0.07 of the voltage applied to the outer coil. The voltage-current characteristics are linear about up to 2.4 $A_{RMS}$ for the outer coil, 2.3 $A_{RMS}$ for the inner coil, and 2.7 $A_{RMS}$ for two coils. When these values are exceeded, a gradual increase of the impedance is observed.

Fig. 4 presents AC losses caused by currents in various coils (detailed study of AC losses in melt cast BSCCO cylinders has been published early in [12,13] and other works). Comparison of the AC losses in the considered designs shows that the losses in the two connected coils are substantially less than those caused by each coil (with doubled turn number). At a low current (< 2.4 $A_{RMS}$) AC losses caused by the inner coil are less and increase faster than the losses due to a current in the outer coil. This result is well described by expressions (6) and (8). The inverse correlation observed at a higher current can be explained by the nonuniform distribution of the current in the cylinder wall.

According to expressions (6) and (8) one should expect that the losses $Q_z$ in the case when the coils are connected in series are 4 times less than the losses $Q_1$ caused by a current in a single coil. Fig. 4 shows that $Q_z/Q_1$ is about 2 at a current above 2$A_{RMS}$. The main cause of this discrepancy is the difference of the actual voltage-current characteristic of a BSCCO superconductor from the step-wise characteristic used in Bean's model. This is confirmed by the numerical calculations of the losses in a superconducting slab subjected to a sinusoidal magnetic field parallel to the slab surface (Fig. 5) where the *E-J* characteristic of the slab has the form $E = E_0 \left( J/J_0 \right)^n$ with the index $n = 7$, and the critical current density depends on the magnetic field according (14) (the Kim-Anderson relation). One can see that for these characteristics, the losses are obtained about proportional to the magnetic field squared while Bean's model gives cubic dependence of the losses on the field. The pronounced difference of the actual AC losses in HTS samples from the values predicted by Bean's model was shown in [12, 13] where the heating effect was also taken into account.

The analysis of the AC losses obtained in our present experiment shows that they are well fitted by power-law dependence on current. As an example, Fig. 6 presents (in the logarithmic scale) the AC losses in the cylinder vs. the current in the outer coil. The point of intersection of two linear fittings corresponds to onset of the increase in the winding impedance. Both the change of the index in the power law and the increase of the impedance are associated with the complete magnetic flux penetration as well as with the growth of the cylinder temperature due to losses. This is illustrated by Fig. 7 where the magnetic flux in the magnetic core is presented as a function of the current in the outer coil for two cases. In the first case, 2 $A_{RMS}$, the amplitude of the current induced in the cylinder is less than the critical value and the flux is small. In the case of 3.4 $A_{RMS}$ when the amplitude of this current is higher than the critical value, the magnetic flux increases disproportionally with current: for considered cases the current increases by 1.7 times at the same time the flux grows tenfold at $i = 0$.

The typical oscilloscope traces of the voltage drop across the model and current in the protected circuit under a fault event are presented in Fig. 8. In spite of the fact that a fault current is limited about tenfold, the characteristic voltage blips associated with the transition into the normal state (see for example [4]) do not appear and, what is more, the cylinder is practically not heated during a fault event. The last is concluded from independence of the voltage-current loops on time (Fig. 8b). Such regime of the current limitation was observed early and explained by the power-law voltage current characteristic and relatively high heat capacity of the cylinder [14].

To determine the activation current, we propose the curves of the current induced in the cylinder vs. the winding current (Fig. 9). The cylinder current $i_{cyl}(t)$ can be calculated using Eqs. (3) and (7).

At low currents, the magnetic flux produced by the cylinder current compensates the flux produced by the winding current and the model impedance is low. The dependence $i(i_{cyl})$ is almost linear (Fig. 9a); a small deviation from linearity is explained by losses in the cylinder and non-ideal magnetic coupling of the cylinder and winding. At a higher current in the winding, the loop expands and the curve deviates from the linear dependence (Fig. 9b). The activation current can be determined as a point where the deviation starts (points A, Fig. 9b, c) or as a half of the difference between the currents in points B and C. Both approaches give close values of the activation current: for the design with outer winding this current is about 6.5 A (Fig. 9b) whereas for two connected coils it is larger than 7.5 A (Fig. 9c) (the value reduced to 200 turns in every coil).

Note that the activation currents obtained from experiments of a sudden fault are noticeably higher than the currents of the complete penetration and the quenching currents measured under step-by-step increasing current (compare Figs. 2, 4, and 9). The possible reason of this behavior is heating of the cylinder due to AC losses.

## 4. Conclusion

The division of the primary winding of an SC cylinder based FCL into two sections, the first of which is placed inside the cylinder and the second is outside the cylinder, decreases the device impedance under the nominal regime and AC losses in the cylinder. As a result of reducing AC losses, the rated and activation currents of the FCL are increased.

The experimental results confirm these conclusions. The experiments carried out with the cylinder characterized by power law voltage-current dependence showed that under a steady-state regime the current of the magnetic field penetration through the cylinder wall is less than the quenching current which, in its turn, is less than the activation current under a sudden fault. The main cause of this is heating of a superconductor due to AC losses.

The obtained results can be generalized to more complicated SC devices such as current-limiting transformers [5,7] or current-limiting transformer with energy redistribution [17], where several windings are magnetically coupled with a SC cylinder. One or more windings are located on the same leg that the cylinder, others are on neighboring legs of the magnetic system. The windings located on other legs produce the magnetic flux inside the cylinder. To reduce AC losses and increase the activation current, it is recommended to divide the winding located on the cylinder leg to two sections so that to provide the equal magnetic fields at the inner and outer surfaces of the cylinder.

**Acknowledgements**
This research was supported by Grants from the Ministry of Infrastructure of Israel, the Ministry of Science, Culture & Sport of Israel and the Russian Foundation for Basic Research, the Russian Federation.

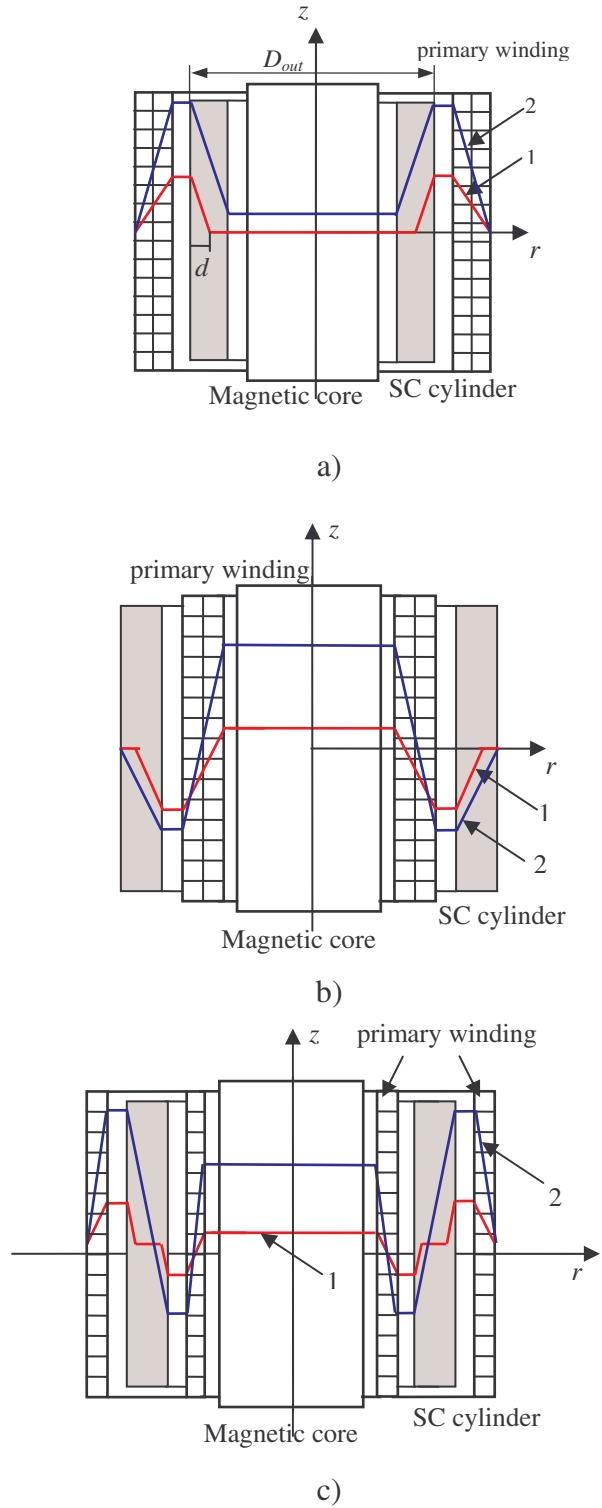

Fig. 1. (Color online) Mutual location of the SC cylinder and primary winding in FCL: SC cylinder inside the primary winding (a), outside the winding (b) and between two sections (c). Red lines 1 show the magnetic field distributions when the winding current is below the activation value and current induced in the cylinder less than the critical value. Blue lines 2 show the distributions when the winding current is above the activation value.

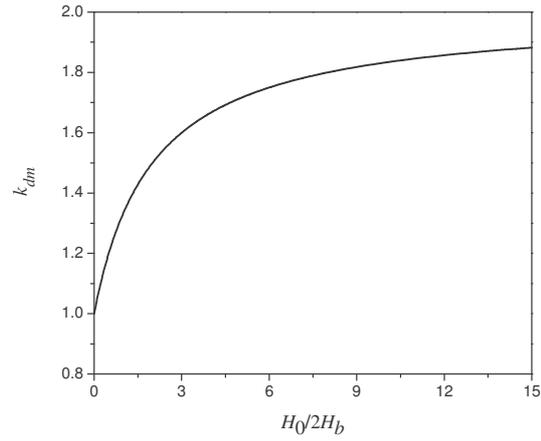

Fig. 2. Ratio of the maximum penetration depth in the device with a single winding and in the device with a two-sectional winding vs. $H_0/2H_b$.

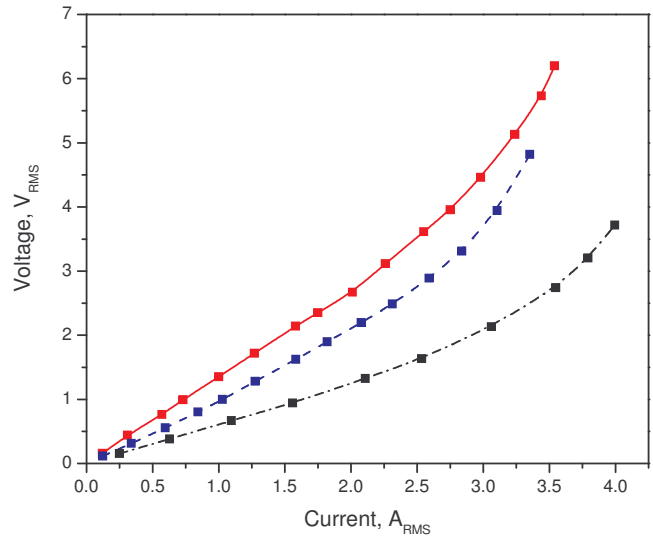

Fig. 3 (Color online) Voltage-current characteristics of the model: red solid line – winding is placed outside the cylinder, blue dashed line – winding is inside the cylinder, black dash-dotted line – two windings connected in series (the values of current and voltage are reduced to the total number of turns of two windings equaled 400).

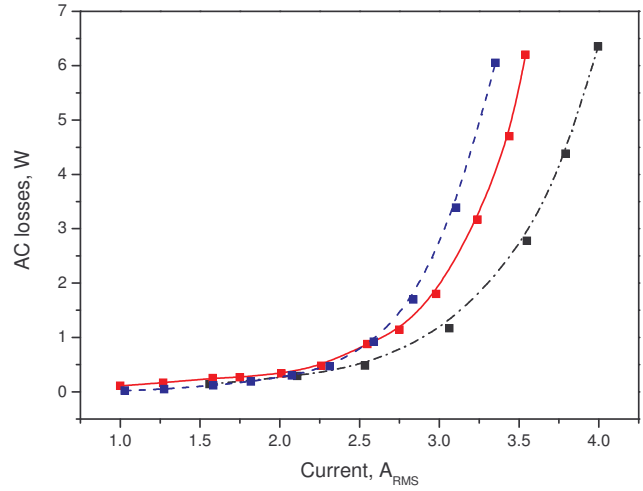

Fig. 4 (Color online) AC losses in the cylinder vs. current in the primary winding: red solid line – winding is placed outside the cylinder; blue dashed line – winding is inside cylinder; black dash-dotted line – two windings connected in series (current is reduced to the total number of turns of two windings equaled 400).

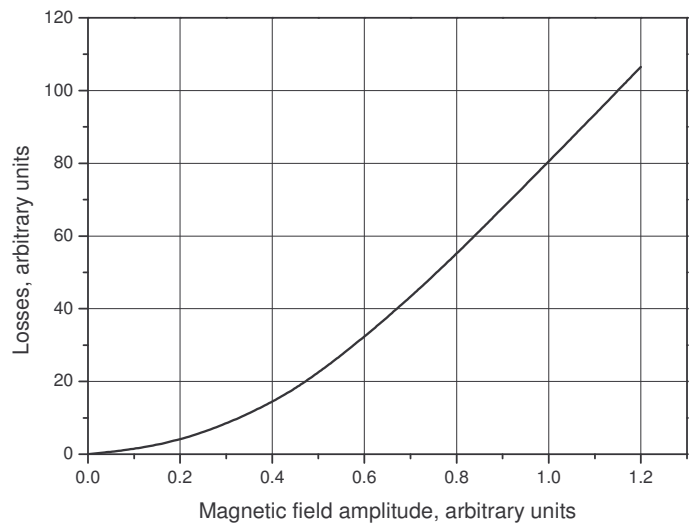

Fig. 5 Losses calculated for a superconducting slab in a sinusoidal magnetic field parallel to the slab surface. The calculations are based on the power-law characteristic $E = E_0 \left( J/J_0 \right)^n$, $n = 7$, and Kim-Anderson's dependence of the critical current density on magnetic field.

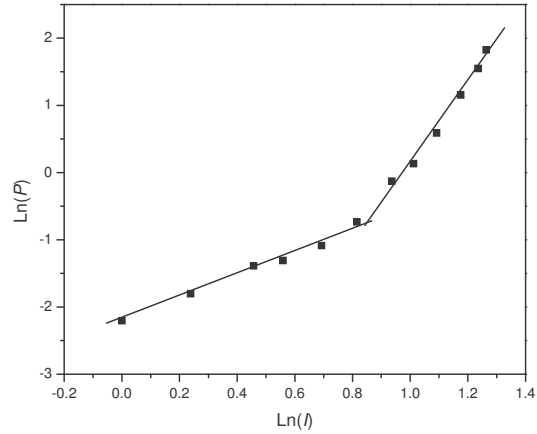

Fig. 6 AC losses in the cylinder vs. current in the outer winding (logarithmic scale).

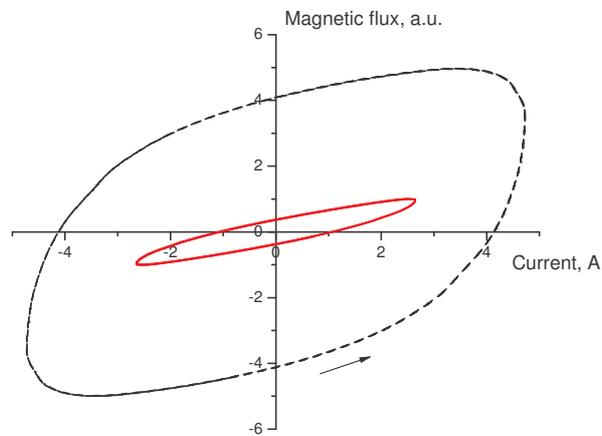

Fig. 7 (Color online) Magnetic flux (arbitrary units) inside the cylinder vs. current in the outer coil; red solid line – at 2 $A_{RMS}$, black dashed line – 3.4 $A_{RMS}$. The arrow shows the direction of the bypass. The magnetic flux was obtained as an integral of the voltage drop across the inner coil through time.

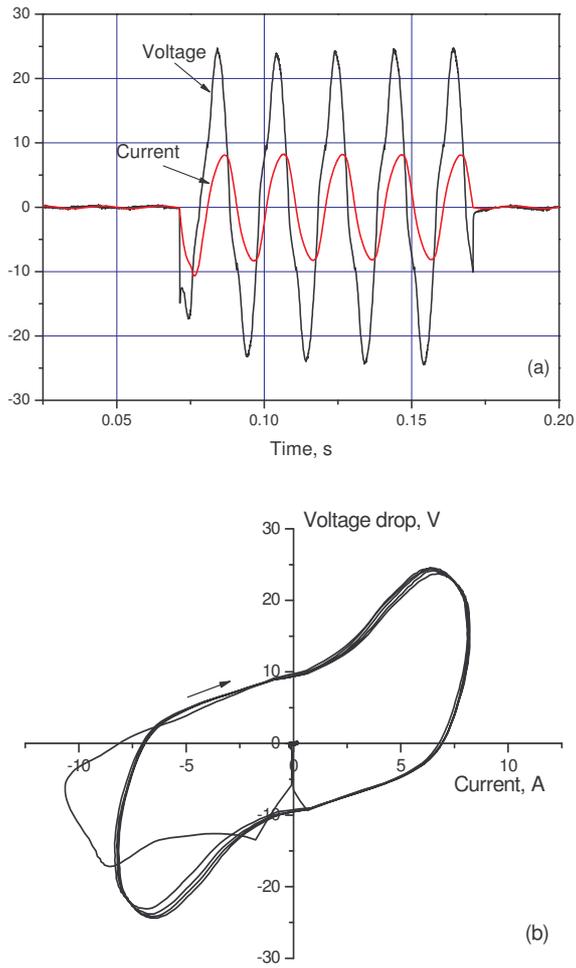

Fig. 8 (Color online) Oscilloscope traces (a) of current in the circuit (red) and voltage drop across the model (black) and dependence of the voltage drop on the current in the winding (b). The arrow shows the direction of the bypass. The results are presented here for the outer coil.

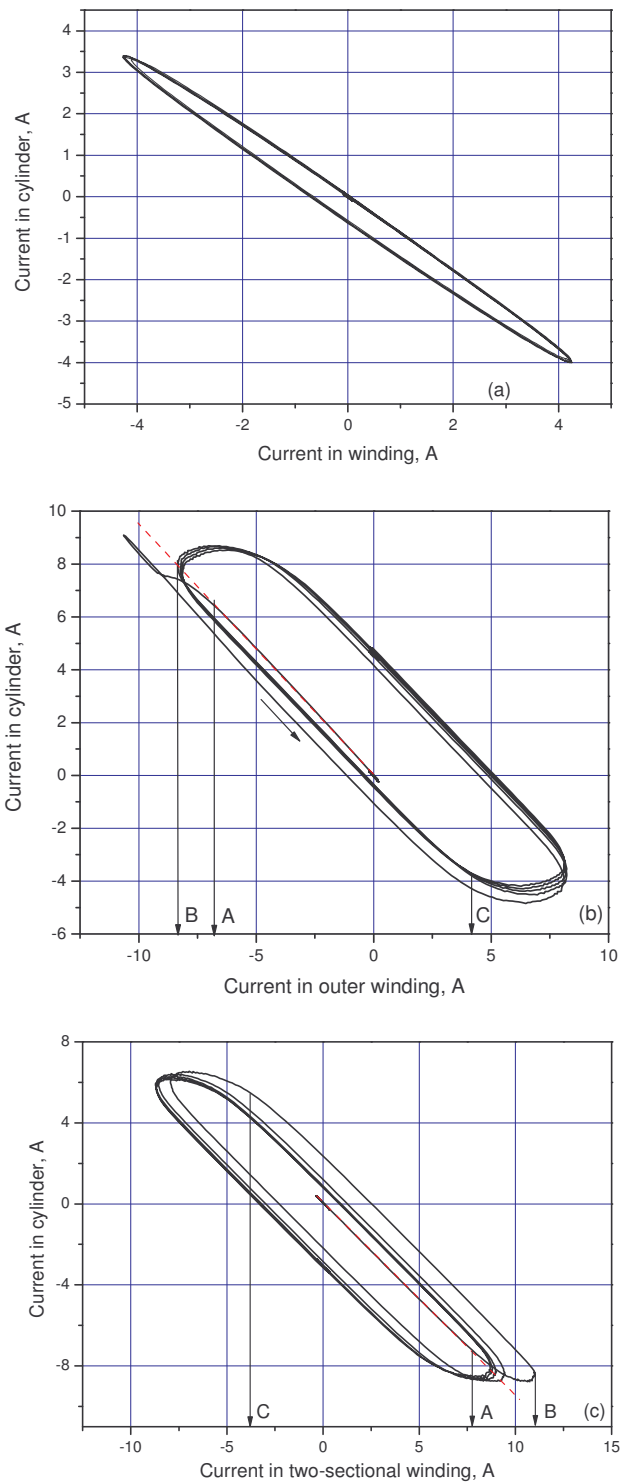

Fig. 9 Current in the cylinder vs. current in the winding: (a) outer winding – the current amplitude is less than the activation value; (b) outer winding – the amplitude is higher than this value; (c) two windings connected in series (current in the two-sectional winding is reduced to the total number of turns equaled 400). The current in the cylinder is reduced to primary winding.